\documentclass[runningheads]{llncs}
\usepackage[T1]{fontenc}
\usepackage{graphicx}

\usepackage[bookmarksdepth=1,bookmarksopen]{hyperref}

\usepackage{listings}
\usepackage{tikz}
\usetikzlibrary{calc,automata,positioning,shapes.geometric}
\usepackage{venndiagram}

\usepackage{wrapfig}
\usepackage{paralist}
\usepackage{colortbl}
\usepackage{algorithm}
\usepackage[compatible,noend]{algpseudocode}

\usepackage[group-digits=integer, group-minimum-digits=4,
                list-final-separator={, and }, 
                free-standing-units, unit-optional-argument, 
                detect-weight=true, 
                per-mode=fraction, 
                ]{siunitx}
								
\usepackage{booktabs}
\usepackage{dcolumn}
\newcolumntype{d}[1]{D{.}{.}{#1} }
\newcolumntype{s}[1]{D{/}{/}{#1} }

\usepackage{relsize}
\usepackage{xspace}
\newcommand\tool[1]{{{\smaller\scshape #1}\xspace}}
\newcommand\definetool[2]{\newcommand{#1}{\tool{#2}\xspace}}

\definetool{\cpachecker}{CPAchecker}
\definetool{\uautomizer}{Ulti\-mate Auto\-mizer}
\definetool{\uautomizerSV}{SV25-UA}
\definetool{\uautomizerNWA}{SEQ-UA}
\definetool{\uautomizerONE}{PAR-1-UA}
\definetool{\uautomizerTWO}{PAR-2-UA}
\definetool{\uautomizerFOUR}{PAR-4-UA}
\definetool{\uautomizerSIX}{PAR-6-UA}
\definetool{\uautomizerPL}{PAR-(2|4|6)-UA}
\definetool{\DSS}{DSS}
\definetool{\benchexec}{BenchExec}
\definetool{\smtinterpol}{SMTInterpol}
\definetool{\mathsat}{MathSAT5}
\definetool{\cvcfour}{CVC4}

\usepackage{varwidth}
\usepackage{tcolorbox}
\newcommand{\llbox}[1]{
		\vspace*{2pt plus 3pt minus 1pt}
		\begin{tcolorbox}[width=\columnwidth, colframe=black, boxrule=0.25mm, top=1mm, left=1mm, right=1mm, bottom=1mm]
			#1
		\end{tcolorbox}
}

\newcommand{\multiLineComment}[1]{}
\makeatletter
\RequirePackage{hyperref}%

\def\orcidID#1{{\href{https://orcid.org/#1}{\protect\raisebox{-1.25pt}{\protect\includegraphics{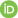}}}}}
\makeatother

\def\BibTeX{{\rm B\kern-.05em{\sc i\kern-.025em b}\kern-.08em
    T\kern-.1667em\lower.7ex\hbox{E}\kern-.125emX}}

\begin{document}

\title{Multi-Threaded Software Model Checking via Parallel Trace Abstraction Refinement}

\titlerunning{Parallel Trace Abstraction Refinement}

%
\author{Max Barth\orcidID{0009-0002-7716-3898}
\and
Marie-Christine Jakobs\orcidID{0000-0002-5890-4673} }
\authorrunning{M.~Barth \and M.-C.~Jakobs}
\institute{LMU Munich, Germany}

\maketitle              
\begin{abstract}
Automatic software verification is a valuable means for software quality assurance.
However, automatic verification and in particular software model checking can be  time-consuming, which hinders their  practical applicability e.g., the use in continuous integration. 
One solution to address the issue is to reduce the response time of the verification procedure by leveraging today's multi-core CPUs.

In this paper, we propose a solution to parallelize trace abstraction, an abstraction-based approach to software model checking. 
The underlying idea of our approach is to parallelize the abstraction refinement.
More concretely, our approach analyzes  different traces (syntactic program paths) that could violate the safety property in parallel.
We realize our parallelized version of trace abstraction in the verification tool~ \uautomizer and perform a thorough evaluation.
Our evaluation shows that our parallelization is more effective than sequential trace abstraction and can provide results significantly faster on many time-consuming tasks. 
Also, our approach is more effective than \DSS, a recent parallel approach to abstraction-based software model checking.


\end{abstract}

\section{Introduction}
Software failures can have severe consequences.
In particular for safety-critical systems, it is therefore important to not only detect software failures but also prove their absence.
While software testing may reveal failures, it typically fails to prove their absence.
In contrast, formal software verification techniques support both, the detection of specification violations (i.e., bugs) and the proof that a program is correct wrt.\ its specification.
However, formal verification needs to become less time-consuming to fit into the software development process.

One solution is to leverage today's multi-core CPUs and reduce the response time via parallelization.
Parallel portfolio approaches, e.g.,~\cite{DBLP:conf/icse/DwyerEPP07,DBLP:conf/kbse/HolzmannJG08,DBLP:conf/vmcai/ChakiK16,DBLP:conf/fase/BeyerKR22}, which run different software verifiers in parallel, facilitate easy parallelization.
Unfortunately, they also waste a lot of resources because they only use the result of the first verifier that finishes successfully.
Other parallelization approaches, e.g.~\cite{BAM-parallel,RangedProgramAnalysis,DBLP:journals/pacmse/0001K024}, split the verification task (program plus property) into several subtasks and analyze them in parallel.
Still, finding a good splitting is difficult.

In this paper, we follow an orthogonal approach and aim to achieve the reduction in response time by parallelizing the verification algorithm itself.
Many approaches in this category, e.g., \cite{DBLP:conf/spin/GaravelMS01,DBLP:conf/atva/EvangelistaKP13,DBLP:conf/vmcai/LopesR11,DBLP:conf/cc/RodriguezL11,DBLP:conf/sigsoft/SunXZQW0W0LLPG23}, parallelize the computation of the state space, which requires complicated synchronization.
In contrast, we pursue a similar idea as Yin, Dong, Liu, and Wang~\cite{ParallelRefinementSCAR} and aim to parallelize abstraction refinement. 
While Yin, Dong, Liu, and Wang target bounded model checking for concurrent programs, we consider abstraction-based software model checking, more concretely trace abstraction~\cite{DBLP:conf/sas/HeizmannHP09,DBLP:conf/cav/HeizmannHP13}. 

Trace abstraction~\cite{DBLP:conf/sas/HeizmannHP09,DBLP:conf/cav/HeizmannHP13} is an automata-based software model checking approach for safety properties like assertions.
It maintains an automaton (the abstraction), which describes those syntactical program paths that lead to a property violation and whose feasibility still needs to be analyzed.
During verification, trace abstraction alternates between (a)~selecting a path from the current abstraction, (b)~checking the feasibility of the selected path, and when proven infeasible (c)~determining an automaton that describes the infeasible path and paths with similar reasons of infeasibility before (d)~refining the abstraction, i.e., removing the determined paths from the abstraction.

We observe that step~(b) and (c) can be done independently when given the selected path and the program.
Hence, our parallelization uses several worker instances for step~(b) and (c), which all consider different paths.
In addition, a single coordinator is responsible for steps~(a) and (d) as well as the communication with the workers.

\begin{figure}[t]

\hfill
\begin{minipage}{0.16\textwidth}
\begin{verbatim}
0: if x>0
1:    x=-x;
2: else
3:    while x>-10
4:       x=x-1;
5: assert x!=0;
\end{verbatim}
\end{minipage}
\hfill
\begin{minipage}{0.400\textwidth}
\begin{tikzpicture}[>=stealth]
 \node (s){};
 \node (l0) [draw, circle,minimum size= 0.5cm , inner sep=0cm, right of = s, node distance=0.75cm, fill=green!20] {$\ell_0$};
 \node (l1) [draw, circle,minimum size= 0.5cm , inner sep=0cm, right  of = l0, node distance=1.5cm, fill=green!20] {$\ell_1$};
 \node (l2) [draw, circle,minimum size= 0.5cm , inner sep=0cm, below   of = l0, node distance=1.5cm, fill=green!20] {$\ell_3$};
 \node (l4) [draw, circle,minimum size= 0.5cm , inner sep=0cm, right of = l1, node distance=1.5cm, fill=green!20] {$\ell_5$};
 \node (l3) [draw, circle,minimum size= 0.5cm , inner sep=0cm, below   of = l4, node distance=1.5cm, fill=green!20] {$\ell_4$};
 \node (l5) [draw, circle,minimum size= 0.5cm , inner sep=0cm, right of = l4, node distance=1.5cm, fill=green!20] {$\ell_6$};
 \node (le) [draw, circle,minimum size= 0.5cm , inner sep=0cm, below  of = l5, node distance=1.5cm, fill=green!20,accepting] {$\ell_{e}$};

 \draw[->] (s) -- (l0);
\draw[->] (l0) to node[above , font=\footnotesize] {\texttt{x>0}} (l1);
\draw[->] (l0) to node[left , font=\footnotesize] {!(\texttt{x>0})} (l2);
\draw[->] (l1) to node[above,font=\footnotesize] {\texttt{x=-x;}} (l4);
\draw[->,bend left=10] (l2) to node[above,font=\footnotesize] {\texttt{x>-10}} (l3);
\draw[->, bend left=10] (l3) to node[below ,font=\footnotesize] {\texttt{x=x-1;}} (l2);
\draw[->] (l2) to node[above ,font=\footnotesize,rotate=25] {!(\texttt{x>-10})} (l4);
\draw[->] (l4) to node[above,font=\footnotesize, ] {\texttt{x!=0}} (l5);
\draw[->] (l4) to node[above,font=\footnotesize,rotate=-45] {!(\texttt{x!=0})} (le);
\end{tikzpicture}
\end{minipage}
\hfill
\caption{Source code (left) and program automaton (right) of our example  \texttt{NotZero}}
\label{fig:example}
\end{figure}
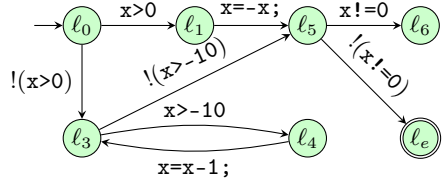

We demonstrate our parallelization for two worker threads on the program\footnote{For more details on the program automaton, we refer to the next section.} shown in Fig.~\ref{fig:example}. 
First, the coordinator selects the two paths~\mbox{\(\pi_1=x>0,\)} \(x=-x,!(x!=0)\) and \(\pi_2=!(x>0),!(x>-10),!(x!=0)\), one following the if and the other the else branch.
Note that both paths are semantically infeasible. 
The coordinator assign paths \(\pi_1\) and \(\pi_2\)  to separate workers.
Assume that the coordinator first gets the result for path~\(\pi_1\).
Thus, it excludes \(\pi_1\) from its current abstraction.
For demonstration purposes, we assume that after the abstraction refinement the result for path~\(\pi_2\) is not yet available.
Hence, the coordinator selects a third path~\(\pi_3=!(x>0),x>-10,x=x-1;,!(x>-10),\) \mbox{\(!(x!=0)\)} and assigns it to the free worker.
Next, we assume that the coordinator receives the result for \(\pi_2\) and gets to know that it can exclude all paths following the else branch.
After removing the paths from the else branch from the abstraction, the coordinator detects that there are no further paths to check and reports the program is proven safe.
While this example wastes CPU~time on processing path~\(\pi_3\) , the response time of the verification should still be shorter because it processes paths~\(\pi_1\) and \(\pi_2\) in parallel. 

We implement the demonstrated idea of parallel trace abstraction refinement in the tool \uautomizer and thoroughly evaluate it on verification tasks of the SVBenchmark.
Our evaluation shows that parallelization can be beneficial wrt.\ effectiveness and response time.

In summary, our paper makes the following contributions.
\begin{compactitem}
	\item We conceptually develop a parallel version of the trace abstraction algorithm (Sec.~\ref{sec:parTraceAbstraction}), an abstraction-based algorithm for software model checking targeting safety properties, and integrate it into the tool \uautomizer. 
	\item We perform a rigorous experimental evaluation of parallel trace abstraction (Sec.~\ref{sec:eval}) considering \num{14560}~C~verification tasks of the SVBenchmark and four different numbers of workers. 
	Also, we compare against 
	a state-of-the-art parallelization approach. 
\end{compactitem}

\section{Sequential Trace Abstraction}
In this paper, we aim to parallelize trace abstraction~\cite{DBLP:conf/sas/HeizmannHP09,DBLP:conf/cav/HeizmannHP13}, a software verification technique to analyze whether a program adheres to a given safety property. 
Trace abstraction, which is also known as automata-based software model checking, maintains and iteratively refines (i.e., reduces the size of) a set of traces, a subset of the syntactical program paths that may cause a property violation. 
During verification, the set of traces forms the current (trace) abstraction of the program and is represented via an automaton. 
In the following, we formally introduce the notion of traces, define the initial trace abstraction, and explain the sequential verification procedure that iteratively refines trace abstractions.

\subsection{Traces and Initial Trace Abstraction}
Our goal is to use traces to represent (syntactic) program paths.
To this end, a trace describes the order in which a (syntactic) program path performs its operations.
In our presentation, we consider two types of program operations: assignments and assume statements (Boolean expressions), which both operate on a set~\(V\) of integer variables\footnote{Our implementation supports C statements.}.
The set~\(\Sigma\) describes the set of all possible program operations.
Hence, \emph{traces} are sequences~\(\pi\in\Sigma^*\) of program operations.
We order these traces based on their prefix relation.
More concretely, we write \(\pi'\preceq\pi\) if there exists \(\pi''\in \Sigma^*\) such that \(\pi=\pi'\pi''\).


During verification, we consider sets~\(\mathcal{S}_\pi\) of traces that represent syntactic program paths that will violate the safety property of interest if they are semantically feasible.
For the  efficient representation of a set~\(\mathcal{S}_\pi\) of traces, we use a finite automaton~\({A}\) with alphabet~\(\Sigma\).
The automaton's language is exactly the set of traces, i.e., \(\mathcal{L}({A})=\mathcal{S}_\pi\).
The verification starts with the set~\(\mathcal{S}_\mathcal{I}\) of all traces that model a syntactic program path that potentially violates the safety property of interest.
Since one may encode safety properties as the unreachability of certain program locations (named error locations)~\cite{DBLP:conf/cav/HenzingerJMNSW02}, we restrict ourselves to safety properties encoded by error locations.
Thus, the set~\(\mathcal{S}_\mathcal{I}\) contains all traces that model a syntactic program path that may end in an error location.

To describe the set~\(\mathcal{S}_\mathcal{I}\), 
we need to model the program as an automaton that accepts the traces from the set~\(\mathcal{S}_\mathcal{I}\).
To this end, we use 
a \emph{(program) auto\-maton}~\(A_\mathcal{P}=(L, \delta_\mathcal{P}, \ell_0, F_\mathcal{P})\)~\cite{DBLP:conf/sas/HeizmannHP09}.
Its set~\(L\) of states represents the program locations, which are closely related to the program counter values. 
The initial location~\(\ell_0\in L\) relates to the program counter value at the start of the program.
Furthermore, the transition relation~\(\delta_\mathcal{P}\subseteq L\times\Sigma\times L\) models the program's control-flow. 
Transitions~$(\ell,\texttt{op}, \ell') \in \delta_\mathcal{P}$ describe which operations~$\texttt{op}$ 
may be executed at a given program location~$\ell$ and where to proceed ($\ell'$) after the execution of operation~$\texttt{op}$.
Finally, the set~\(F_\mathcal{P}\subseteq L\) contains the accepting states, which correspond to the program's error locations and determine the safety property. 
Hence, a program automaton accepts exactly the traces in \(\mathcal{S}_\mathcal{I}\).

Figure~\ref{fig:example} shows the source code (left) and the resulting program automaton (right) of our example program~\texttt{NotZero}.
Except for the error location~\(\ell_e\), the index of a location in the program automaton refers to the corresponding line number in the source code of the program.
Furthermore, the automaton contains one edge per assignment and two assume statement edges per if, while, or assert statement, namely one for each evaluation of their condition. 
To let the violation of the assertion result in a property violation, the false evaluation of the assertion's condition leads to the error location~\(\ell_e\).


\subsection{Iterative Trace Abstraction Refinement}
Next, we describe the iterative verification procedure, namely the sequential trace abstraction algorithm.
The verification procedure employs counterexample-guided abstraction refinement (CEGAR)~\cite{DBLP:conf/cav/ClarkeGJLV00} and is shown in Alg.~\ref{alg:modelChecking}.

  \begin{algorithm}[t]
		\caption{Sequential trace abstraction procedure~\cite{DBLP:conf/sas/HeizmannHP09}}\label{alg:modelChecking}
  \begin{algorithmic}[1]
  \REQUIRE program automaton~\(A_\mathcal{P}\) 
  \STATE $A = A_{\mathcal{P}}$;
  \WHILE{True}  
  \IF{$\mathcal{L}(A)$ == $\emptyset$}
		return SAFE;
  \ENDIF

   \STATE select $\pi\in\mathcal{L}(A)$
   \STATE (result, interpolants) = checkSAT($\varphi_{\pi}$);
   \IF{result == SAT}
				return UNSAFE;
    \ENDIF
        
    \STATE  $A_\pi$ = generateAutomaton($\pi$, interpolants, $A$);
     \STATE $A = A \setminus A_\pi$
  \ENDWHILE
 
  \end{algorithmic}  
  \end{algorithm}

\begin{figure}[t]
\centering
\begin{tikzpicture}[>=stealth]
     \node (s){};
     \node (l0) [draw, circle,minimum size= 0.5cm , inner sep=0cm, right of = s, node distance=0.5cm, fill=green!20] {$q_0$};
     \node at ($(l0.north)+(0,0.025)$) [anchor=south,font=\footnotesize] {\{true\}};
     \node (l1) [draw, circle,minimum size= 0.5cm , inner sep=0cm, right  of = l0, node distance=1.75cm, fill=green!20] {$q_3$};
     \node at ($(l1.north)+(0,0.025)$) [anchor=south,font=\footnotesize] {\{$x\leq0$\}};
     \node (l4) [draw, circle,minimum size= 0.5cm , inner sep=0cm, right of = l1, node distance=2cm, fill=green!20] {$q_5$};
     \node at ($(l4.north)+(0,0.025)$) [anchor=south,font=\footnotesize] {\{$x\leq-10$\}};
     \node (le) [draw, circle,minimum size= 0.5cm , inner sep=0cm, right  of = l4, node distance=1.75cm, fill=green!20,accepting] {$q_{e}$};
     \node at ($(le.north)+(0,0.025)$) [anchor=south,font=\footnotesize] {\{false\}};
    \draw[->] (s) -- (l0);
    \draw[->] (l0) to node[below, font=\footnotesize]{!\texttt{(x>0)}} (l1);
    \draw[->] (l1) to node[below,font=\footnotesize] {!\texttt{(x>-10)}} (l4);
    \draw[->, loop below] (l1) to node[below,font=\footnotesize] {$\texttt{x>-10},x=x-1;$} (l1);
    \draw[->] (l4) to node[below,font=\footnotesize] {!(\texttt{x!=0})} (le);
\end{tikzpicture}

\caption{An interpolant automaton for trace \(!(x>0),!(x>-10),!(x!=0)\)}
\label{fig:interpolant-automaton}
\end{figure}
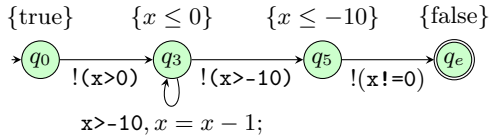

The algorithm gets a program automaton, which accepts the traces that represent syntactic program paths that lead to a property violation (i.e., an error location).
During verification, the algorithm maintains automaton~\({A}\) to represent the current trace abstraction, which contains all traces that model a syntactic program path to an error location for which the semantic feasibility has not yet been checked.
Line~1 initializes the trace abstraction with the program automaton~\(A_\mathcal{P}\).
Lines~2--8 perform the iterative refinement of the trace abstraction.
The verification will end with result SAFE (i.e., property proven) in line~3 if the language of the automaton~\({A}\) is empty, i.e., there does not exist a trace that models a feasible program path that ends in an error location.
If the language is non-empty, there still exist syntactic program paths to error locations for which the semantic feasibility has not yet been checked.
Hence, line~4 uses breadth-first search to select a trace $\pi\in\mathcal{L}(A)$ representing one such path.
Thereafter, line~5 checks the semantic feasibility of that path.
To this end, it first constructs an SSA-based formula encoding $\varphi_{\pi}$ of $\pi$ and subsequently checks \(\varphi_\pi\)'s satisfiability with an SMT solver.
For example, for trace \(!(x>0),!(x>-10),!(x!=0)\) we check formula~\(\neg(x_0>0)\wedge \neg(x_0>-10)\wedge\neg(x_0\neq 0)\) and for trace~\mbox{\(x>0,x=-x;,\)} \(!(x!=0)\) we check formula~\(x_0>0\wedge x_1=-x_0\wedge\neg(x_1\neq 0)\).
If the result of the satisfiability check is SAT, we proved that the syntactic program path represented by trace~\(\pi\) is semantically feasible.
Since all program paths represented by a trace \(\pi\in\mathcal{L}(A)\) end in an error location, we found a feasible counterexample and line~6 returns UNSAFE (i.e., property violation detected).
Otherwise, we refine the trace abstraction in lines~7 and 8.
First, we construct an \emph{interpolant automaton}~$A_\pi$~\cite{DBLP:conf/sas/HeizmannHP09} that accepts $\pi$ and other traces with similar reasons of semantic infeasibility.
Note that the construction of $A_\pi$  uses abstraction~\(A\) to focus on traces that are still relevant and, thus, to make the construction more efficient.
Figure~\ref{fig:interpolant-automaton} shows a possibly interpolant automaton for trace~\(!(x>0),\) \mbox{\(!(x>-10),!(x!=0)\)}.
The automaton encodes the trace and extends it with an additional loop.
Also, the automaton records the infeasibility argument for its accepted traces by annotating its states with assertions.
Typically, we use interpolation to derive the assertions from the unsatisfiability proof of the trace's formula encoding.
Given the interpolant automaton, which by construction only accepts traces that are semantically infeasible, line~8 refines our current trace abstraction.
More concretely, it removes the traces accepted by the interpolant automaton from the current trace abstraction by computing the difference of our current abstraction (automaton~\(A\)) and the interpolant automaton~$A_\pi$. 

\multiLineComment{
\begin{example}
  \label{ex:updatedAbstraction}
\center{
\begin{tikzpicture}[>=stealth]
  \node (s){};
  \node (l0) [draw, circle,minimum size= 0.5cm , inner sep=0cm, right of = s, node distance=0.5cm, fill=green!20] {$\ell_0$};
  \node (l1) [draw, circle,minimum size= 0.5cm , inner sep=0cm, below right  of = l0, node distance=1.25cm, fill=green!20] {$\ell_1$};
  \node (l4) [draw, circle,minimum size= 0.5cm , inner sep=0cm, below of = l1, node distance=1.5cm, fill=green!20] {$\ell_4$};
  \node (le) [draw, circle,minimum size= 0.5cm , inner sep=0cm, below  of = l4, node distance=1.25cm, fill=green!20,accepting] {$\ell_{e}$};
  \node (l5) [draw, circle,minimum size= 0.5cm , inner sep=0cm, left of = le, node distance=1.25cm, fill=green!20] {$\ell_5$};
 
 \draw[->] (s) -- (l0);
 \draw[->] (l0) to node[right, font=\footnotesize] {\texttt{x>0;}} (l1);
 \draw[->] (l1) to node[right,font=\footnotesize] {\texttt{x=-x;}} (l4);
 \draw[->] (l4) to node[above,font=\footnotesize, rotate= 45] {\texttt{x != 0}} (l5);
 \draw[->] (l4) to node[right,font=\footnotesize] {!(\texttt{x != 0})} (le);
 \end{tikzpicture}
}
\end{example}
}

\multiLineComment{

\subsection{Interpolation Automata}\mcjFB{why is this a separate subsection?}
\mcjFBIn{Are these details really needed for this paper, could a much shorter version be integrated with the previous subsection?}

In each CEGAR iteration we aim to exclude as many spurious counterexamples from the set of considered counterexamples as possible.
Therefore, we want to construct an \emph{interpolant automaton}~\cite{DBLP:conf/sas/HeizmannHP09}
 that accepts more than just the spurious counterexample $\pi$.
We achieve this in two steps. 
First we determine the reason for infeasibility, then we generalize the automaton 
such that more traces that are infeasible for the same reason are accepted.
Given an infeasible error trace e.g.: $$\pi_i = !(x > 0); !(x > -10); !(x != 0);$$ we obtain the reason for its infeasibility via interpolation.

In our approach, we have multiple interpolation techniques available \cite{DBLP:conf/tacas/HeizmannCDGHLNM18} for example: 
Craig Interpolation~\cite{DBLP:journals/jsyml/Craig57a, DBLP:conf/tacas/JhalaM06,DBLP:conf/cav/McMillan06}, Tree interpolation \cite{mcmillan2013solving, DBLP:conf/popl/HeizmannHP10}and Strongest Post Condition and Weakest Pre Condition utilizing unsat cores and quantifier elimination \cite{DBLP:conf/sigsoft/DietschHMNP17}.
In this paper we will not go into any further detail on interpolation and refer to the cited literature instead.

Once we have the interpolant sequence of a trace, we take the straightline automaton for trace $\pi$ and generalize it~\cite{DBLP:conf/sas/HeizmannHP09}.
In practice, this is a complicated and time consuming procedure.
In its core, it constitutes to the following:
We utilize the interpolant sequence to annotate the states of the straightline automaton with first order logic formulas 
that represent the values of every variable in the program. 
This annotation aggregates to a Floyd-Hoare-style-like proof of infeasibility.
Then, we generalize the automaton by adding additional transitions while no new variable values are allowed in our data states .
In other words, we ensure the Floyd-Hoare-style-like proof remains intact during generalization, as shown in \ref{ex:generalized}.



\begin{example}
  \label{ex:generalized}
Consider the error trace $\pi \in \mathcal{L}(A_{\mathcal{P}})$ of example program \texttt{NotZero}~\ref{fig:example}.
$$\pi = !(x > 0); !(x > -10); !(x != 0);$$ 
and the sequence of interpolants: \texttt{true}, $x \leq 0$, $x \leq -10$, \texttt{false}
A possible generalized interpolant automaton $A_\pi$ is:
\center{
\begin{tikzpicture}[>=stealth]
     \node (s){};
     \node (l0) [draw, circle,minimum size= 0.5cm , inner sep=0cm, right of = s, node distance=0.5cm, fill=green!20] {$q_0$};
     \node at ($(l0.north)+(0,0.125)$) [anchor=south,font=\footnotesize] {\{true\}};
     \node (l1) [draw, circle,minimum size= 0.5cm , inner sep=0cm, right  of = l0, node distance=1.75cm, fill=green!20] {$q_1$};
     \node at ($(l1.north)+(0,0.125)$) [anchor=south,font=\footnotesize] {\{$x \leq 0$\}};
     \node (l4) [draw, circle,minimum size= 0.5cm , inner sep=0cm, right of = l1, node distance=1.75cm, fill=green!20] {$q_4$};
     \node at ($(l4.north)+(0,0.125)$) [anchor=south,font=\footnotesize] {\{$x \leq -10$\}};
     \node (le) [draw, circle,minimum size= 0.5cm , inner sep=0cm, right  of = l4, node distance=1.75cm, fill=green!20,accepting] {$q_{e}$};
     \node at ($(le.north)+(0,0.125)$) [anchor=south,font=\footnotesize] {\{false\}};
    \draw[->] (s) -- (l0);
    \draw[->] (l0) to node[above, font=\footnotesize]{!\texttt{(x>0)}} (l1);
    \draw[->] (l1) to node[above,font=\footnotesize] {!\texttt{(x>-10)}} (l4);
    \draw[->, loop below] (l1) to node[below,font=\footnotesize] {$\{x=x-1;, \texttt{x > -10} \}$} (l1);
    \draw[->] (l4) to node[above,font=\footnotesize] {!(\texttt{x!=0})} (le);
\end{tikzpicture}
}

We generalized the automaton by adding the self loop at location $q_1$.
This is possible since the two Hoare triple: $\{x \leq 0\} x=x-1; \{x \leq 0\}$ and $\{x \leq 0\} \texttt{x > -10}  \{x \leq 0\}$ are both valid.
\end{example}

}

\section{Towards Parallel Trace Abstraction}\label{sec:parTraceAbstraction}
Next, we describe how we parallelize the iterative trace abstraction procedure,  which we described in the previous section.
The idea is to parallelize the loop performing the iterative refinement of the trace abstraction.

\subsection{Preliminary Considerations}
Looking at the loop (lines 2--8 of Alg.~\ref{alg:modelChecking}), we observe that the only dependency between loop iterations is the trace abstraction represented by automaton~\(A\).
The update of the abstraction in line~8 aggregates the results of the different iterations and therefore, should not be done in parallel.
Nevertheless, the aggregation itself is commutative, i.e., for the abstraction itself it does not matter in which order we use the interpolant automata to refine it. 
Furthermore, we notice that the check in line~5, which inspects the feasibility of trace~\(\pi\), is independent of abstraction~\(A\).
In addition, when generating the interpolant automaton we only use the abstraction to make the generation more efficient.
Hence, using an older, out-of-date abstraction, which accepts more traces, is sound but may not be as efficient as using the latest one.
For the emptiness check in line~3 and the search of a trace~\(\pi\in\mathcal{L}(A)\), it is also sound to use an out-of-date abstraction.
However, using an out-of-date abstraction may likely result in performing unnecessary work, e.g., more refinement iterations, which one could avoid when using the latest abstraction. 
Further taking into account that we experienced that the feasibility check (line~5) and the interpolant automaton generation (line~7) are the most expensive tasks in each iteration, we decide to consider those two tasks for parallel execution.
Thereby, we concede that different threads may need to maintain memory-expensive out-of-date copies of the abstraction.
This is acceptable because from our experience sequential trace abstraction much more often runs out of time than it runs out of memory.


\begin{algorithm}[t]
  \caption{Coordinator that directs parallel trace abstraction}\label{alg:ParallelCEGARMaster}
\begin{algorithmic}[1]
\REQUIRE program automaton~\(A_\mathcal{P}\) 

\STATE $A$ = $A_\mathcal{P}$; \colorbox{gray!10}{$\mathcal{S}_{\pi_\mathrm{ex}}=\emptyset$;}

\WHILE{True}     
\IF{$\mathcal{L}(A)$ == $\emptyset$}
	return SAFE;
\ENDIF

\WHILE {\colorbox{gray!10}{idleWorkerAvailable() $\wedge$ $\mathcal{L}(A)\setminus\mathcal{S}_{\pi_\mathrm{ex}}\neq\emptyset$}}
\STATE select $\pi\in$\colorbox{gray!10}{$\mathcal{L}(A)\setminus\mathcal{S}_{\pi_\mathrm{ex}}$}
\STATE \colorbox{gray!10}{$\mathcal{S}_{\pi_\mathrm{ex}}=\mathcal{S}_{\pi_\mathrm{ex}}\cup\{\pi\}$;}
\STATE \colorbox{gray!10}{giveTaskToWorker($\pi$, $A$);}
\ENDWHILE

		\STATE \colorbox{gray!10}{waitUntilAtLeastOneWorkerResultAvailable();} 
	\WHILE{\colorbox{gray!10}{workerResultAvailable()}}
	\STATE \colorbox{gray!10}{($A_\pi$,  result) = getWorkerResult();}
		\IF {result == SAT}
			 return UNSAFE;
		\ENDIF
		\STATE $A = A \setminus A_\pi$;
	\ENDWHILE

\ENDWHILE\end{algorithmic}
\end{algorithm}

\subsection{Procedure of Parallel Trace Abstraction}
To realize our parallelization, we use a central coordinator (Alg.~\ref{alg:ParallelCEGARMaster}) in combination with several workers (Alg.~\ref{alg:ParallelCEGARWorker}).
The coordinator directs the verification procedure.
To this end, it runs a modified version of the sequential trace abstraction procedure (Alg.~\ref{alg:modelChecking}).
Differences are highlighted in gray and mainly represent synchronization with workers.
In addition, it maintains a set~$\mathcal{S}_{\pi_\mathrm{ex}}$, which tracks  the traces already assigned for analysis, and  uses it to avoid exploring the same trace~\(\pi\) twice.\footnote{Note that in contrast to sequential trace abstraction, which either stops exploration when detecting that a trace is feasible or removes the infeasible trace from the trace abstraction before selecting the next trace, our parallel trace abstraction selects traces while the analysis of traces selected earlier may not have finished.}
Like the sequential procedure, the coordinator  (Alg.~\ref{alg:ParallelCEGARMaster}) initializes and iteratively updates the trace abstraction (lines~1 and 12).
Also, it checks the stopping criteria of the verification procedure (lines~3 and 11).
To maintain the set~$\mathcal{S}_{\pi_\mathrm{ex}}$,  Alg.~\ref{alg:ParallelCEGARMaster}  initializes it with an empty set of already analyzed traces in line~1 and updates it in line~6 after selecting a trace to analyze.
Furthermore, we adapt the trace selection (line~5) such that it only selects traces that are in the current trace abstraction and have not been assigned for analysis yet, i.e., are not in set~$\mathcal{S}_{\pi_\mathrm{ex}}$.
In contrast to the sequential procedure, Alg.~\ref{alg:ParallelCEGARMaster} distributes the selected traces to the available workers for analysis and relies on the workers' results to update its trace abstraction.

The workers (Alg.~\ref{alg:ParallelCEGARWorker}) process the traces.
They perform the feasibility check of the traces (line~1).
For traces  proven infeasible, they also compute the interpolant automaton (line~3) required for refinement. 
Hence, they perform the steps in line 5 and 7 of the sequential verification procedure (Alg.~\ref{alg:modelChecking}).
In addition, they make the result (the result of the feasibility check and the interpolant automaton) available for the coordinator (line~5).
To ensure that they provide a structurally valid result, they construct an automaton accepting the empty language in line~4 if the trace is feasible, i.e., no interpolant automaton is required.

\begin{algorithm}[t]
	\caption{Worker for trace processing}\label{alg:ParallelCEGARWorker} 
  \begin{algorithmic}[1]
	\REQUIRE trace $\pi$, trace abstraction~$A$ with \(\pi\in\mathcal{L}(A)\)
	\STATE (result, interpolants) = checkSAT($\varphi_\pi$);
	\IF {result = UNSAT}
	  \STATE $A_\pi$ = generateAutomaton($\pi$, interpolants, $A$);
		\ELSE ~\colorbox{gray!10}{$A_\pi = (\{q_0\}, \emptyset, q_0, \emptyset)$;}
	  \ENDIF
	  \STATE \colorbox{gray!10}{publishResult($A_\pi$,  result);}
\end{algorithmic}
\end{algorithm}

To coordinate with the workers, each iteration of the coordinator contains a phase to distribute the work to the workers (lines~4--7) followed by a phase to collect and process the workers' results (lines~8--12). 
Each phase is realized with a while loop.
The loop distributing work stops when no further worker or trace to analyze is available, while the loop considering worker results stops if no result is available.
Furthermore, we avoid busy waiting by explicitly waiting until some result becomes available in line~8. 

Using the coordinator together with one worker basically performs sequential trace abstraction, but distributed on two threads.
When we use more than one worker, we need to take particular care of the trace selection.
To avoid redundant work, we already ensure that a trace is only selected once.
Ideally, we would like to select diverse traces, e.g., traces with different reasons for infeasibility.  
We discuss trace selection in more detail in the next section.

\subsection{Procedure for Diverse Trace Selection}
\label{sec-search}
Now, we discuss our trace selection (i.e., line~5 of Alg.~\ref{alg:ParallelCEGARWorker}).
The selection must ensure that it does not select a trace from $\mathcal{S}_{\pi_\mathrm{ex}}$ and ideally it should select a trace with a new reason for infeasibility.
The more the newly selected trace differs from the previously selected traces~$\mathcal{S}_{\pi_\mathrm{ex}}$, the likelier it will be that it provides a different reason for infeasibility.
With this in mind, we aim to steer our selection such that it selects traces that diverge early  from previously selected traces.


\begin{algorithm}[t]
	\caption{Recursive algorithm~\texttt{search} to select diverse traces from a trace abstraction}\label{alg:search}
  	\begin{algorithmic}[1]
	\REQUIRE trace abstraction $A=(Q, \delta, q_0, F)$, search state~\(q\), selected prefix $\pi$,  relevant set of analyzed traces $\mathcal{S}_{\mathrm{rt}}$
	\IF {$\mathcal{S}_{\mathrm{rt}}=\emptyset$}
	    \IF{acceptingStateReachableFrom($q$, $A$)}
    	    \STATE return $\pi$ $\circ$ shortestPathToAcceptingStateFrom($q$, $A$);
			\ELSE~
					return null;
			\ENDIF
	\ENDIF
	\STATE succ = list($\{(q,\cdot,\cdot)\in\delta\}$);
	\STATE sort(succ, $(q, \texttt{op}, q')\rightarrow|\{\pi'\in \mathcal{S}_{\mathrm{rt}}\mid \pi\circ \texttt{op} \preceq\pi'\}|$);

	\WHILE {!empty(succ)}
	    \STATE $(q, \texttt{op}, q')$ = dequeue(succ);
	\IF {$\pi\circ\texttt{op}\notin \mathcal{S}_{\mathrm{rt}}$}	    
    	    \STATE $\pi_r$ = {search}($A$, $q'$, $\pi\circ\texttt{op}$, $\{\pi'\in \mathcal{S}_{\mathrm{rt}}\mid \pi\circ \texttt{op} \preceq\pi'\}$);

 		      \IF { $\pi_r$ != null }~return $\pi_r;$			
					\ENDIF
		 \ENDIF
	\ENDWHILE
	\STATE return null;

\end{algorithmic}
\end{algorithm}

Algorithm~\ref{alg:search} shows our selection procedure, which implements a recursive search procedure through the abstraction.
The algorithm gets the trace abstraction~$A$ to search through, the state~\(q\) where in the abstraction to continue the search,  the already selected prefix~\(\pi\), and  the  set of relevant analyzed traces~$\mathcal{S}_{\mathrm{rt}}$, i.e., the previously selected traces with prefix~\(\pi\) that are not yet known to be infeasible.
Thus,  line~5 of Alg.~\ref{alg:ParallelCEGARWorker} calls the algorithm with the current trace abstraction, the abstraction's initial state~\(q_0\),  an empty prefix trace~\(\varepsilon\), and set~$\mathcal{S}_{\pi_\mathrm{ex}}\cap\mathcal{L}(A)$ of traces already assigned for analysis but not yet known to be infeasible.

To select traces that diverge early from previously selected traces, Alg.~\ref{alg:search}  prioritizes successor transitions that have been reached less often via the prefix~\(\pi\) currently considered by the search procedure. 
If no traces have been selected yet that use the currently considered prefix~\(\pi\), i.e., $\mathcal{S}_{\mathrm{rt}}$ is empty, lines~1--4 return the shortest trace with prefix~\(\pi\) that is in the language of the abstraction  or null if none exist.
The prioritization itself is realized in lines~5--12 of Alg.~\ref{alg:search}.
First, line~5 builds a list of all transitions that start in the current search state~\(q\).
Next, the algorithm sorts these transitions by how likely they lead to a new trace.
We approximate the likelihood with the number of previously selected traces that are continuations of the trace built from the currently selected prefix~\(\pi\) and the transition's operation.
Thereby, a higher number indicates a smaller likelihood.
Thereafter, the while loop realizes the actual prioritized search.
In each iteration, line~8 removes the  transition with the largest likelihood (smallest number of continuations) that has not yet been explored.
The check in line~9 ensures that we skip the transition whenever the transition's operation together with the currently selected trace results in a previously selected trace.
In all other cases, line~10 recursively searches for a trace not previously selected.
The recursive call searches for a continuation of  prefix trace~\(\pi\) extended with the current transition's operation.
Therefore, it continues the search with the successor~\(q'\) of the currently considered transition and adapts the search to those relevant traces that agree with the new prefix.
If the recursive call detects a new trace (\(\pi_r\neq\textnormal{null}\)), we return it in line~11.
However, if the current transition cannot lead us to a new trace, the next loop iteration tries the next transition.
If we fail to generate a new trace with any available transition outgoing of the current search state~\(q\), we return null in line~12 and backtrack.
Note that Alg.~\ref{alg:search} terminates with a non-null result as long as there exists a trace accepted by trace abstraction~\(A\), which is not in the initial set~{$\mathcal{S}_{\mathrm{rt}}$ of analyzed traces.
Algorithm~\ref{alg:ParallelCEGARWorker} guarantees this whenever it selects a new trace.

\subsection{Implementation}
We integrate our approach for parallel trace abstraction into  \uautomizer~\cite{UltimateAutomizerTool},
a Java-based  software verification tool that already supports sequential trace abstraction.
Following our conceptual approach, 
our implementation of parallel trace abstraction reuses the components of the existing sequential trace abstraction and
 mainly adds the parallelization and adapts the trace search.

To realize the parallelization, we use the \texttt{Runnable} interface to implement Alg.~\ref{alg:ParallelCEGARWorker}, 
which processes a given trace~\(\pi\).
Furthermore, we realize the workers with a fixed thread pool, which is provided as an \texttt{ExecutorService}.
Note that the user can configure the number of threads in the pool via a parameter.
To start new trace processing and to retrieve their results, 
 we combine the \texttt{ExecutorService} instance with a \texttt{CompletionService} instance.
The coordinator uses the \texttt{CompletionService} instance to initiate a new trace processing and to retrieve the results from finished trace processing without busy waiting.

To allow for a better comparison between sequential and parallel trace abstraction, 
our implementation of parallel trace abstraction aims to include all traces~\(\pi\) in its analysis that the sequential trace abstraction analyzes.
To this end, we use a slightly modified search for diverse traces.
In the first iteration of each execution of the loop in lines~4--7 of Alg.~\ref{alg:ParallelCEGARMaster}, 
we first apply breadth-first search, the search strategy of sequential trace abstraction, 
 to detect a trace~\(\pi\) accepted by the current trace abstraction (i.e., \(\pi\in\mathcal{L}(A)\)).
If \(\pi\) has not been found before (i.e., \(\pi\notin\mathcal{S}_{\pi_\mathrm{ex}}\)),
 we distribute \(\pi\) to a worker and continue with the next iterations of the loop, which all perform our proposed search (Alg.~\ref{alg:search}).
Otherwise, we apply our proposed search.
However, we do not compute $\mathcal{S}_{\pi_\mathrm{ex}}\cap\mathcal{L}(A)$ when calling the search but  overapproximate it.
To this end, we remove a trace~\(\pi\) from \(\mathcal{S}_{\pi_\mathrm{ex}}\)  when we process the result of its analysis.
Another improvement we employ is that we limit the search for a trace~\(\pi\) to \num{5} seconds\footnote{This limit proved useful in practice. Our observation is that the search typically finishes within one second.} instead of performing the expensive non-emptiness check in line~4 of Alg.~\ref{alg:ParallelCEGARMaster} and treat a timeout of the search as a failed non-emptiness check.
Note that this is sound because we never use the non-emptiness check to determine the result of the verification.



\section{Evaluation}\label{sec:eval}

The goal of our parallelization is to reduce the response time and to increase the effectiveness of the approach. 
Therefore, our evaluation investigates the effect of our parallelization on response time and solved tasks.
Furthermore, we compare our parallelization with a recent parallelization technique that uses software model checking techniques similar to trace abstraction.



\subsection{Experimental Setting}

\textbf{Tool Configurations.}
In our evaluation, we consider four different configurations of our parallel trace abstraction, namely  \uautomizerONE,  \uautomizerTWO, \uautomizerFOUR and \uautomizerSIX.
They differ in the size of the fixed thread pool.
\uautomizerONE uses one worker thread and describes the sequential baseline of our approach.
The other three configurations use two, four, and six worker threads.  
Since even the sequential configuration \uautomizerONE uses a coordinator and worker thread, we also run \uautomizer with the sequential configuration~\uautomizerNWA used in the latest competition on software verification (SV-COMP25).
For all five configurations, we use \uautomizer version~0.3.0-474959ed\footnote{Note that \uautomizerNWA solves slightly more tasks in this version than in its version submitted to SV-COMP~2025.}. 
Furthermore, we use \DSS,  a recent state-of-the-art technique for parallel model checking, which employs distributed summary synthesis~\cite{DBLP:journals/pacmse/0001K024}.  
For \DSS, we use the version provided in artifact~\cite{DBLP:data/11/BeyerKL24a} accompanying the paper.


\textbf{Verification Tasks.}
For our evaluation, we consider verification tasks from the software verification benchmark\footnote{\url{https://gitlab.com/sosy-lab/benchmarking/sv-benchmarks/-/tree/svcomp25-final}}  (SVBenchmark), which is  used by the international competition on software verification (SV-COMP)~\cite{SVCOMP25}.
A verification task in SV-COMP consists of a program and a property.
We restrict our evaluation to all tasks from SV-COMP~2025  that consider (a)~sequential C~programs, the programs supported by trace abstraction in \uautomizer, and (b)~property~\texttt{unreach-call}, a safety property which states that certain error locations (encoded as calls to the function~\texttt{reach\_error}) must not be reached.
In total, we consider  \num{14560}~tasks, \num{11158} of them fulfill the property while  \num{3402} of them violate the property. 

\textbf{Evaluation Environment.}
The machines, we use, contain \num{33}\,GB of memory and an Intel Xeon E3-1230 v5 CPU with \num{8}~processing units and a frequency of \num{3.40}\,GHz.
Their operating system is a 64-bit  Ubuntu~24.04 with Linux kernel~6.8 and the installed Java version is OpenJDK~21.0.7. 
%
We limit each run of any tool configuration on a verification task  to  \num{8}~CPU cores, 
 \num{15}\,min of wall time and \num{30}\,GB of memory 
and enforce the limits with \benchexec~\cite{DBLP:journals/sttt/BeyerLW19} (version~3.18).

\begin{table}[t]
  \caption{Per tool configuration, number of correctly and incorrectly reported property violations (alarms) and property 
  ad\-herence (proofs)}
  \label{tab:solvedOverview}
\centering
  \begin{tabular}{l d{0} d{0} d{0} d{0} d{0} d{0}}
  \toprule
&   \multicolumn{1}{c}{\uautomizerNWA} &    \multicolumn{1}{c}{ \uautomizerONE} &    \multicolumn{1}{c}{ \uautomizerTWO} &     \multicolumn{1}{c}{\uautomizerFOUR} &     \multicolumn{1}{c}{\uautomizerSIX} &   \multicolumn{1}{c}{\DSS}\\
\midrule
correct alarms & 1128 & 1140 & 1241 & 1243 &  \textbf{1245} & 291\\
correct proofs & 3377 & 3695 & 3747 & \textbf{3860}  & 3858 & 1263\\
incorrect alarms & 0 & 0 & 0 & 0 & 1 & 317\\
incorrect proofs & 0 & 0 & 0 & 0 & 0 & 62\\

  \bottomrule
  \end{tabular}
  \end{table}

\subsection{Impact of Parallelization on Effectiveness}
In this section, we study the impact of our parallelization on effectiveness, i.e., the number of (correctly) solved tasks.
To this end, let us first look at Tab.~\ref{tab:solvedOverview}, which shows for every tool configuration (column) the number of correctly and incorrectly reported alarms (property violations) and proofs (property satisfactions).
When studying the trace abstraction configurations (first five columns), we observe that \uautomizerONE performs better than the SV-COMP 2025 configuration \uautomizerNWA.
One reason is that our new search strategy allows us to recover from exceptions that occur during trace processing.
Due to worse effectiveness, we do not further consider \uautomizerNWA  in the remaining evaluation.
Furthermore, we observe that only \uautomizerSIX reports incorrect results, namely one incorrect alarm.
Studying  the incorrect result, we observe that all other configurations fail to solve this task.
A more detailed manual inspection of the incorrect analysis result lets us believe that the unsoundness is caused by the reused \uautomizer functionality and not our parallelization.
Further studying Tab.~\ref{tab:solvedOverview}, we observe that with increasing number of threads the number of solved alarms and proofs typically increases.
The only exception is  \uautomizerSIX, which solves less proofs than  \uautomizerFOUR.
We investigated this further and observe that \uautomizerONE can correctly solve \num{2219} of \num{4835} by analyzing less than six traces, i.e., \uautomizerSIX does not provide any benefits while causing additional overhead, which may cause timeouts.
When restricting the set of tasks for which \uautomizerONE analyzes at least six traces, i.e., all configurations have its worth, \uautomizerSIX has the largest number of correct proofs.

%
\llbox{
Effectiveness typically increases when the number of workers increases.
}

\begin{figure*}[t]
\centering
 \includegraphics[width=0.32\textwidth]{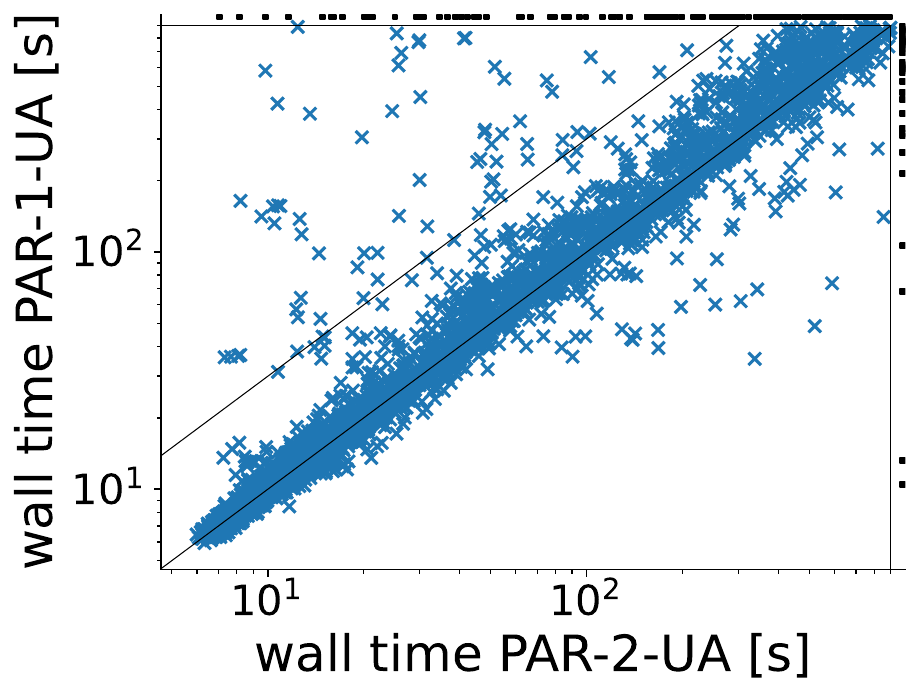}
 \hfill
 \includegraphics[width=0.32\textwidth]{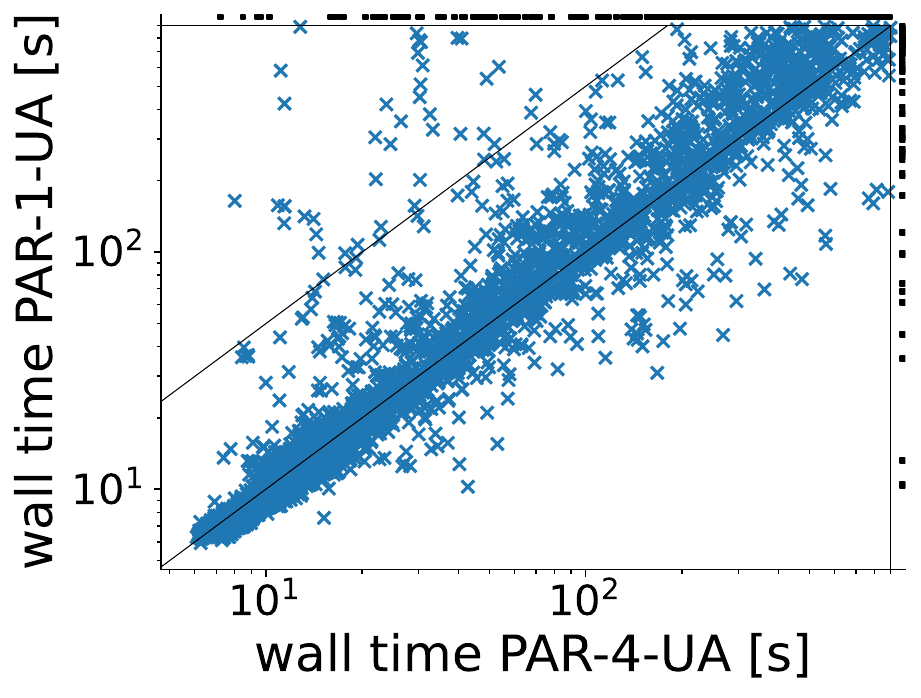}
\hfill
 \includegraphics[width=0.32\textwidth]{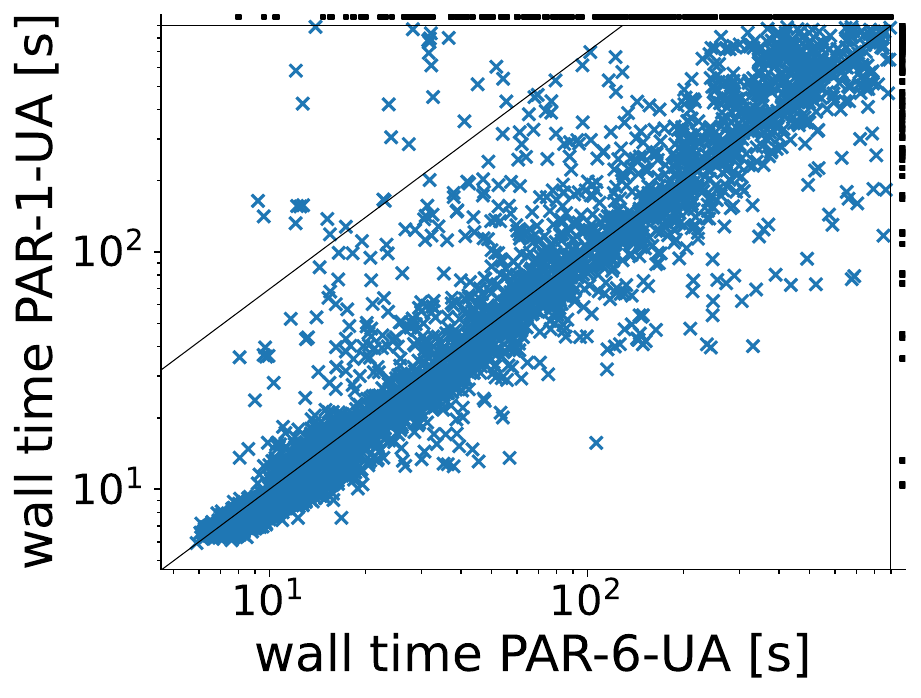}
\caption{Scatter plot that compares the wall time of \uautomizerTWO (left), \uautomizerFOUR (middle), and \uautomizerSIX (right) with the wall time of \uautomizerONE (y-axes) on tasks solved by both configurations. Additionally, points outside the rectangle represent tasks either not solved by \uautomizerONE (top) or \uautomizerTWO, \uautomizerFOUR, and \uautomizerSIX (right)}
\label{fig:scatterWall}
\end{figure*}

\subsection{Effect of Parallelization on Response Time}
Next, we study whether parallelization may reduce the response time (i.e., wall time).
To observe the immediate effect of parallelization on response time, we aim to compare configurations for trace abstraction that only differ in the number of worker threads.
Hence, we compare the wall time of our approach with one worker (\uautomizerONE)
with the configurations \uautomizerTWO, \uautomizerFOUR, and \uautomizerSIX that utilize \num{2}, \num{4}, and \num{6} workers. 
Figure~\ref{fig:scatterWall} shows 
one scatter for each of the parallel configurations
\uautomizerTWO, \uautomizerFOUR, and \uautomizerSIX.
Each scatter plot compares the wall time of the respective parallel configuration  with the wall time of  \uautomizerONE.
Thereby, it contains data points~\((x,y)\) for all tasks correctly solved by at least one of the two compared configurations, 
i.e., \num{5036}~tasks for \uautomizerTWO,  \num{5178}~tasks for \uautomizerFOUR, and  \num{5203}~tasks for \uautomizerSIX.
One point~\((x,y)\) in the scatter plot compares the wall time used by  \uautomizerTWO, \uautomizerFOUR, and \uautomizerSIX, respectively,
with the wall time used by \uautomizerONE (\(y\)) for one particular task.
(Black) points above the top border show tasks that \uautomizerONE does not solve.
Similarly, tasks right of the right border show tasks that the respective configuration using multiple worker threads does not solve.
Depending on the scatter plot, \uautomizerONE fails to solve between \num{201} and \num{368}~tasks, 
while \uautomizerTWO, \uautomizerFOUR, and \uautomizerSIX fail to solve \num{48}, \num{75}, and \num{100}~tasks, respectively.
Sometimes, a configuration with multiple worker threads fails while \uautomizerONE solves the tasks close to the timeout.
Thus, the overhead for parallelization may cause the failure.
In other cases, we observe that the configuration with multiple worker threads analyzes more traces and still fails.
Hence, we believe that the order in which the trace abstraction is refined may differ.
This may lead to different trace abstractions such that the relevant traces that need to be analyzed to succeed will not be selected and analyzed early enough.
Still, the number of failures of \uautomizerONE for which the parallelization succeeds are much larger than vice versa.


Next, let us look at tasks solved by both of the compared configurations, i.e., the points in the rectangle.
All three scatter plots contain points below the diagonal,  i.e., \uautomizerONE responds faster, and above the diagonal, i.e., \uautomizerONE responds slower.
More importantly, there are points in each plot that are above the line parallel to the diagonal.
This line represents the upper bound for the speedup achievable according to Amdahl's law.\footnote{The upper bound of the speedup is identical with the thread number.}
For a point to be above that upper bound, the parallelization needs to skip some work done by \uautomizerONE.
For example, due to parallelization we may find a property violation earlier, find a better loop abstraction earlier, or in general get to a smaller abstraction earlier, which may
allow us to skip processing some traces considered by \uautomizerONE.
In detail, \uautomizerTWO responds faster for \num{49}\% (\num{2369} of  \num{4787}) of the commonly solved tasks
and achieves a speedup larger than \num{3} for \num{67}~tasks.
Furthermore, \uautomizerFOUR  responds faster  for \num{43}\% (\num{2002} of  \num{4760}) of the commonly solved tasks
and achieves a speedup larger than \num{5} for \num{42}~tasks.
\uautomizerSIX  responds faster \num{34}\% (\num{1619} of  \num{4735})  of the commonly solved tasks
and achieves a speedup larger than \num{7} for \num{31}~tasks.
%
Moreover, we observe that if parallelization is slower,  it will typically not take that much longer.
Indeed, it only takes between \num{0.68} and \num{1.99} additional seconds in the median and between \num{9.58} and \num{14.59} additional seconds on average.
When not profiting from parallelization, we do not lose much.

  \begin{figure}[t]
  \centering
\hfill
   \includegraphics[width=0.32\textwidth]{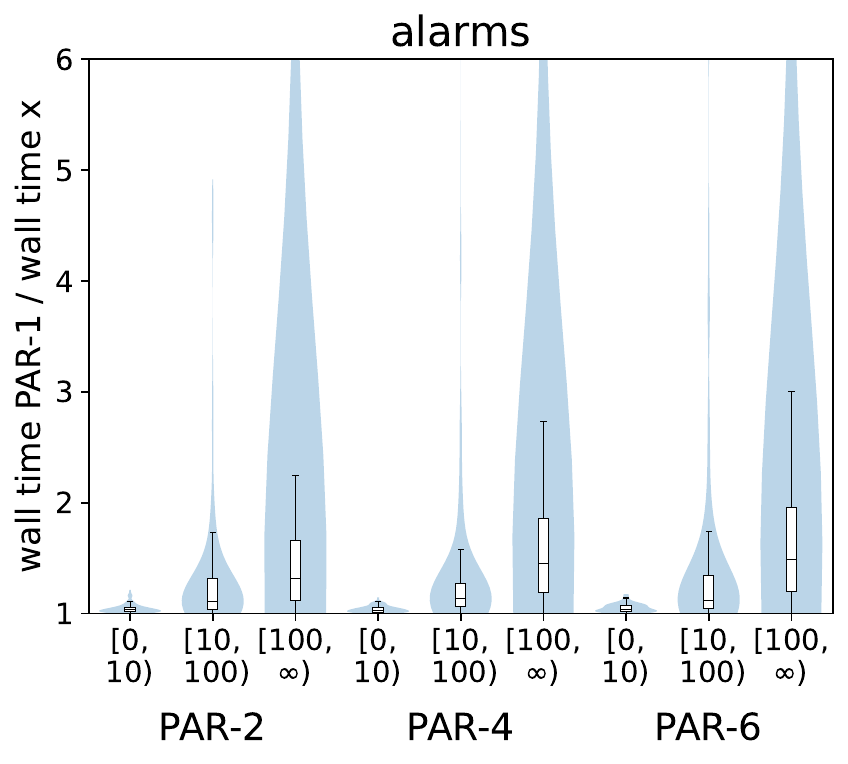}
  \hfill
   \includegraphics[width=0.32\textwidth]{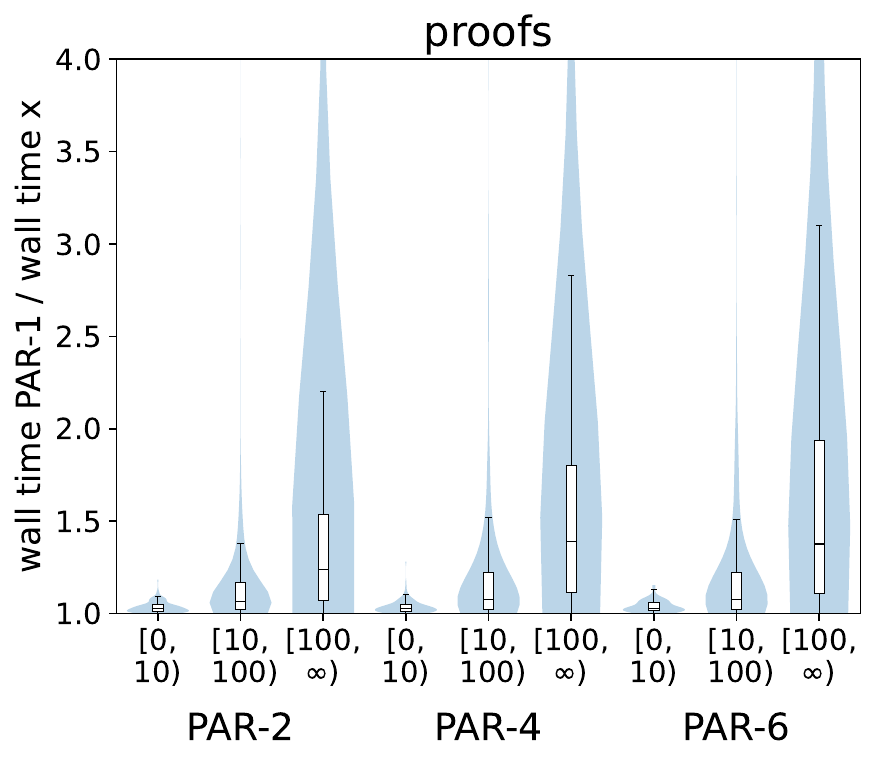}
\hfill
  \caption{Speed up on tasks for which \uautomizerONE responds more slowly. 
  Grouped by the wall time of \uautomizerONE into the intervals $[0,10)$, $[10, 100)$,  and $[100,\infty)$. 
  On the left, we only consider tasks with property violations (alarms) while on the right we only consider tasks that adhere to the specification (proofs).}
  \label{fig:violinSpeed}
  \end{figure}

Next, let us study those tasks for which \uautomizerTWO, \uautomizerFOUR, and \uautomizerSIX respond faster, respectively.
Our goal is to investigate whether the benefit of parallelization is different for proof and alarm detection as well as for more difficult tasks, i.e., tasks for which \uautomizerONE takes longer.
Figure~\ref{fig:violinSpeed} shows two plots, 
the left considers alarms 
and the right considers proofs. 
Each plot contains three combined violin and box plots per configuration using multiple worker threads considering tasks solved by both, the respective configuration with multiple worker threads and \uautomizerONE.
However, the combined plots only consider those commonly solved tasks for which \uautomizerONE is slower wrt.\ wall time and show the distribution of the speedup 
(i.e., wall time of  \uautomizerONE divided by wall time of the other configuration) for tasks which \uautomizerONE solves in less than \num{10}\s\, of wall time, in \num{10}-\num{100}\s\, of wall time, and more than \num{100}{\s} of wall time.
For each group, we observe that from left to right the violins and boxes become larger and the whiskers become higher,
 i.e., we may be able to save more wall time when the verification is complex and takes longer.
However, the variance for more complex tasks is also higher.
Furthermore, we observe that the variance increases when using more worker threads.
It seems that with more worker threads the possible speedup depends more and more on the benchmark.
When comparing the left and right plots in Fig.~\ref{fig:violinSpeed}, 
we observe that the violins on the left are larger.
It seems easier to achieve high speedups from parallelization when detecting property violations.

\llbox{
Parallelization can significantly reduce response time, particularly for time-consuming tasks, while exhibiting negligible negative effects when no improvements are achieved.
}

\subsection{Comparison Against Existing Parallelization Techniques}
%
Finally, we compare our parallelization to state-of-the-art parallelization approaches.
We are aware of two recent, relevant parallelization approaches: 
ranged program analysis~\cite{RangedProgramAnalysis} 
and distributed summary synthesis (\DSS)~\cite{DBLP:journals/pacmse/0001K024}.
Like us, both may use a verification procedure based on predicates, 
CEGAR, and interpolation.
Since we technically failed to port ranged program analysis to Ubuntu~24.04 (the OS environment for our experiments),
 we only compare our approach to \DSS. 
Since \DSS makes use of all \num{8}~available cores, we decide to compare \DSS to our configuration~\uautomizerSIX, which uses the most threads.


\textbf{Effectiveness.}
First, we compare the number of solved tasks. 
To this end, we first compare the columns of Tab.~\ref{tab:solvedOverview} that are labeled with \uautomizerSIX and \DSS.
We observe that  \DSS correctly detects significantly fewer alarms and proofs while reporting much more incorrect results, both alarms and proofs.
One reason for the higher number of incorrect results is \DSS's limited support for pointers and arrays, which is known to cause incorrect results.


\begin{wrapfigure}{r}{0.4\textwidth}
\vspace{-0.5\baselineskip}
 \includegraphics[width=0.32\textwidth]{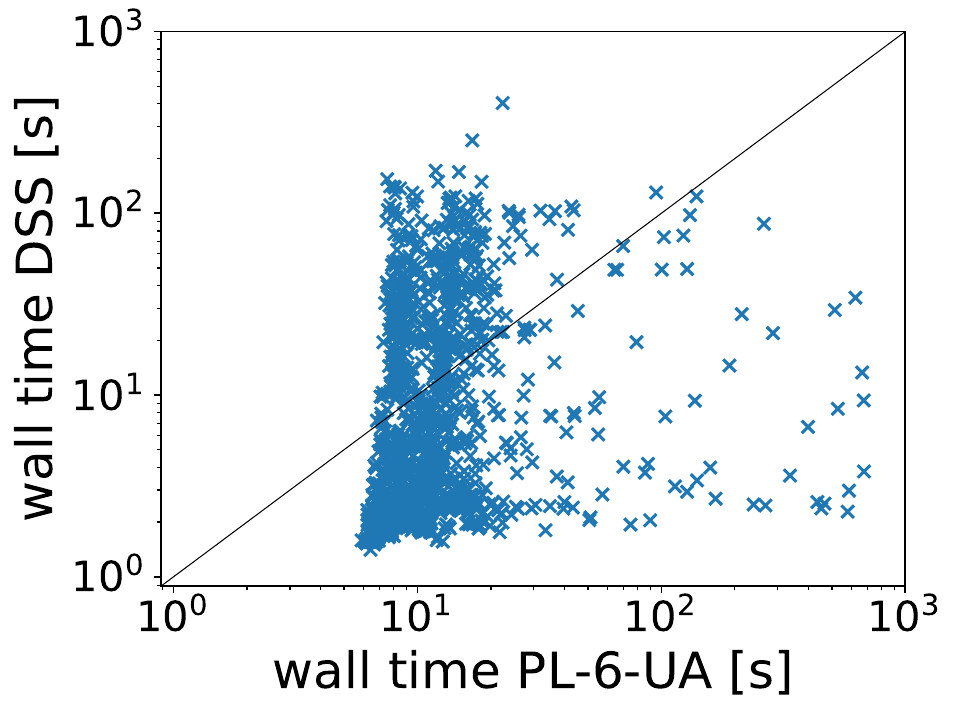}
\caption{Comparing the wall time of our approach~ \uautomizerSIX (x-axis) and competitor \DSS (y-axis)}
\label{fig:compEfficiency}
\vspace{-0.5\baselineskip}
\end{wrapfigure}
\textbf{Efficiency.}
To compare the efficiency in terms of response time (wall time), 
 we restrict our comparison to the \num{1395} tasks correctly solved by our approach (i.e.,  \uautomizerSIX) and \DSS.
Figure~\ref{fig:compEfficiency}  shows a scatter plot 
that contains one \((x,y)\) for each of \num{1395} tasks that compares the wall time of our approach (\(x\)) with the wall time of \DSS (\(y\)).
We observe that there exist points above and below the diagonal, 
i.e., our approach solves the task faster 
than \DSS and vice versa.
A detailed analysis reveals that \DSS uses less time for \num{66}\% (\num{915} of \num{1395}) of the tasks. 
For these \num{915} tasks,  \DSS is faster by \num{7.1}\s\, in the median. 
A large proportion of the \num{7.1}\s\, are caused by the larger startup time of \uautomizer and, thus,  of \uautomizerSIX, which is about \num{4.7}\s\, larger.

\llbox{
Our approach is more effective and regularly more efficient than \DSS. 
}
\subsection{Threats to Validity}
Our implementation of the parallelization may contain bugs.
We think it is unlikely since we only observed one incorrect result when using six worker threads. 
%
Still, our results may not generalize.
Despite using all sequential C tasks from SV-COMP that consider the \texttt{unreach-call} safety property, 
which encode different safety properties, the set may not reflect all real-world sequential programs and safety properties.
Also, results may differ with other resource limits.

\section{Related Work}
We are aware of one other software verification approach~\cite{ParallelRefinementSCAR} that parallelizes the CEGAR refinement step.
The approach performs scheduling constraint based abstraction refinement, a verification approach for concurrent programs, which is based on bounded model checking (BMC).
Its parallelization runs  multiple instances of the verification procedure in parallel and lets each instance use a random search strategy parameterized by the process ID  to detect different counterexamples (i.e., syntactic paths to property violations).
The refinement results (i.e., scheduling constraints) are exchanged via a shared data structure.
In contrast, we consider a different verification technique, only parallelize the analysis of the counterexample but perform a more sophisticated search. 

Instead of parallelizing the refinement steps, Lopes and Rybalchenko~\cite{DBLP:conf/vmcai/LopesR11} parallelize the exploration of the abstract state space in predicate-abstract model checking.
Also, there are approaches~\cite{DBLP:conf/spin/GaravelMS01,DBLP:journals/entcs/Holzmann08,DBLP:conf/atva/EvangelistaKP13} that parallelize the state-space construction in explicit model checking.
Parallelized state-space construction may even use GPUs~\cite{DBLP:conf/mochart/EdelkampS10,DBLP:conf/spin/BartocciDS14,DBLP:conf/tacas/WijsB14}.
In context of static analysis, e.g., dataflow analyses, some approaches parallelize the computation of abstract facts~\cite{DBLP:conf/cc/RodriguezL11,DBLP:conf/cgo/DeweyKH15,DBLP:journals/pacmpl/KimVT20,DBLP:conf/sigsoft/SunXZQW0W0LLPG23}
while in LTL model checking, several approaches~\cite{DBLP:conf/fsttcs/BrimCKP01,DBLP:journals/fmsd/BarnatC06,DBLP:conf/atva/EvangelistaPY11} parallelize the emptiness check of  the language described by the product of system automaton and negated specification.
Also, some approaches run complete algorithmic blocks of an analysis in parallel like  base- and step-case in k-induction~\cite{DBLP:journals/corr/abs-1111-0372} or invariant generation for k-induction~\cite{DBLP:conf/cav/0001DW15}.
Furthermore, verifiers may use implementations of decision diagrams~\cite{DBLP:books/sp/18/DijkP18} or SAT and SMT solvers~\cite{DBLP:books/sp/18/HyvarinenW18} that can operate in parallel, e.g., use multiple threads to perform an operation.

Parallelizing the verification algorithm itself may be complex.
Thus, another class of approaches splits the verification task into several subtasks and analyzes them in parallel.
One option is to split the set of program paths, which is e.g., applied to dataflow analyses~\cite{DBLP:conf/tacas/ShermanD18}, BMC~\cite{DBLP:conf/dac/GanaiG08,DBLP:conf/kbse/NguyenS0TP17,DBLP:conf/ppopp/InversoT20}, software model checking~\cite{RangedProgramAnalysis,RangedProgramAnalysis-journal,RangedProgramAnalysisInstrumentation,DBLP:conf/icse/RichterCJW25}, and symbolic execution~\cite{Sym-Range,Sym-SSP,SynergiSE}.
Another strategy is to decompose the program into code blocks, as e.g., done in dataflow analyses~\cite{DBLP:conf/icse/ShiZ20,DBLP:journals/jss/StievenartEPR21}, verifiers like Bolt~\cite{DBLP:conf/pldi/AlbarghouthiKNR12}, Coverity~\cite{DBLP:conf/sigsoft/McPeakGR13}, BAM~\cite{BAM-parallel}, or \DSS~\cite{DBLP:journals/pacmse/0001K024}, and static application security testing~\cite{DBLP:conf/sigsoft/ChristakisCFLMP22}.
When one needs to analyze multiple properties, e.g., different assertions, one can also analyze the individual properties in parallel~\cite{DBLP:journals/sigsoft/YangDW15,DBLP:conf/ifm/KumarBLDUB16,DBLP:conf/fmcad/MarescottiGHS17}.

Even simpler than splitting the verification task is it to run different verifiers or different verifier configurations in parallel.
Approaches like PredatorHP~\cite{DBLP:conf/tacas/PeringerSV20}, Nacpa~\cite{Nacpa}, or CoVeriTeam's parallel portfolio~\cite{DBLP:conf/fase/BeyerKR22} run different approaches in parallel but in isolation.
In contrast, some approaches~\cite{DBLP:conf/fmics/AbrahamSBFH06,DBLP:conf/vmcai/ChakiK16,DBLP:conf/cav/ChampionMST16,DBLP:conf/cav/GacekBWWG18,DBLP:conf/vmcai/BlichaHMS20} that run multiple SAT or SMT based verification instances in parallel may share information, e.g., perform lemma sharing.
Finally, swarm verification~\cite{DBLP:conf/kbse/HolzmannJG08,DBLP:journals/tse/HolzmannJG11} and parallel randomized state-space search~\cite{DBLP:conf/icse/DwyerEPP07} aim to diversify the search through the state space, e.g., by letting various instances use different search orders. 

While it is not straightforward to combine our approach with other algorithmic parallelization techniques, splitting the verification task or running different verifier instances on a verification task is orthogonal to our parallelization approach and can easily be combined with our approach.

\section{Conclusion}
Trace abstraction is an automata-based software verification technique,  which checks safety properties of software. 
Its original verification algorithm implemented in \uautomizer has not been designed to work in parallel.
To let trace abstraction benefit from modern compute hardware, in particular multi-threading, we therefore propose a parallelization approach for trace abstraction.
Our approach parallelizes the abstraction refinement, i.e., our approach analyzes multiple, distinct syntactic paths, which lead to an error location, in parallel.
Thereby, each analysis checks the feasibility of the respective path and in case of infeasibility computes an interpolant automaton to refine the trace abstraction.
A coordinator uses a specialized search procedure to find diverse paths that lead to an error location, distributes the found paths to available workers, and applies the refinement to the global trace abstraction.
Since our parallelization is orthogonal to parallel portfolio approaches, which run multiple verifiers in parallel, or parallelization approaches that split the verification task, we can freely combine it with those approaches.

We evaluated the implementation of our parallel trace abstraction in \uautomizer on a large set of C~verification tasks from the SVBenchmark.
Thereby, we observe that parallelization increases the effectiveness, i.e., typically, we solve more tasks with more workers.
Nevertheless, parallelization will only make sense if the number of traces analyzed by  sequential trace abstraction is at least as large as the number of workers.
Furthermore, we observe that in particular for longer running tasks parallelization may reduce the wall time significantly.
For several tasks, the speedup for the wall time is significantly larger than the upper bound given by Amdahl's law for task parallelization, i.e., parallel trace abstraction may allow us to skip some of the verification steps performed by sequential trace abstraction.
This further confirms the value of parallel trace abstraction.


%
%
%
 \bibliographystyle{splncs04}
 \bibliography{bibtex}

\end{document}